\documentclass[12pt]{iopart}

\usepackage{iopams}
\usepackage{url,graphicx}
\begin{document}

\title{Constructing quantum circuits for maximally entangled multi-qubit states using the genetic algorithm}

\author{Zheyong Fan$^{1}$\footnote{Email: brucenju@gmail.com},
        Hugo de Garis$^{2}$,
        Ben Goertzel$^{2,3}$,
        Zhongzhou Ren$^{1}$, and
        Huabi Zeng$^{1}$}

\address{$^1$ Department of Physics, Nanjing University, Nanjing, 210093, China}
\address{$^2$ Artificial Intelligence Institute, Computer Science
              Department, Xiamen University, Xiamen, Fujian Province, China}
\address{$^3$ Novamente LLC}

\begin{abstract}
Numerical optimization methods such as hillclimbing and simulated annealing have been applied to search for highly entangled multi-qubit states.   Here the genetic algorithm is applied to this optimization problem -- to search not only for highly entangled states, but also for the corresponding quantum circuits creating these states.  Simple quantum circuits for maximally (highly) entangled states are discovered for 3, 4, 5, and 6-qubit systems; and extension of the method to systems with more qubits is discussed.  Among other results we have found explicit quantum circuits for maximally entangled 5 and 6-qubit circuits, with only 8 and 13 quantum gates respectively. One significant advantage of our method over previous ones is that it allows very simple construction of quantum circuits based on the quantum states found.
\end{abstract}

\maketitle

\section{Introduction}

\textit{Quantum entanglement} \cite{bengtsson06, horodecki09}, enabling states with correlations that have no classical analogue, is one of the central concepts differentiating quantum information from classical information.  Entanglement is essential to many quantum information protocols such as quantum key distribution, dense coding, and  quantum teleportation, and is also thought to play important roles in quantum computational speedup \cite{jozsa03}.

Abstractly, a  bipartite \emph{pure} state is called \emph{entangled} if it is not decomposable to a tensor product of two states of the two subsystems. Quantitatively, there are measures \cite{plenio07} which tell us how entangled a state is.  Maximally or highly entangled states are of particular interest because their entanglement provides a valuable resource which can be used to perform tasks that are otherwise difficult or impossible.

Maximally entangled quantum states of small systems are well known. For 2-qubit states, the Bell state
$|\textmd{Bell}\rangle = (|00\rangle + |11\rangle)/\sqrt{2}$
is maximally entangled on all counts: maximal entanglement entropy,  maximal violation of the Bell inequality, and complete mixture of its one-party reduced states.  The 3-qubit generalization of the Bell state is the GHZ state
$|\textmd{GHZ}3\rangle
= (|000\rangle + |111\rangle)/\sqrt{2}$,
which is also maximally entangled.

One can easily generalize the GHZ state to general $n$-qubit states,  $|\textmd{GHZ}n\rangle = (|00\cdots 0\rangle \pm |11\cdots 1\rangle)/\sqrt{2}$.
However, for 4 or more qubits, these states are not maximally entangled \cite{higuchi00}, and in fact display below-average entanglement, as shown by the numerical calculations of Borras {\it et al} \cite{borras07}.

Since the mathematical structures of multi-qubit states are complex, numerical optimization methods have been found very helpful in the search for maximally or highly entangled states.  Using \emph{negativity} \cite{zyczkowski98,vidal02,kendon02} as the entanglement measure,
Brown {\it et al}  \cite{brown05} performed a numerical search for highly entangled states of 2, 3, 4, and 5 qubits using the hillclimbing optimization method. Although it searches through a space including \emph{mixed} states, their method ultimately converges to \emph{pure} states. They have successfully found a \emph{simple} (in a technical sense of \emph{simple} to be given below) maximally entangled 5-qubit pure state, which has application to quantum teleportation, superdense coding, and quantum state sharing \cite{muralidharan08}.

There are many entanglement measures besides the negativity \cite{plenio07}. Borras {\it et al} \cite{borras07} presented a detailed comparison of the results obtained by different search procedures based upon different entanglement measures. They searched only pure states and discovered a simple maximally entangled 6-qubit state, the utility of which for quantum teleportation and quantum state sharing has been explored \cite{choudhury09}.

Motivated by the simplicity of these maximally entangled 5-qubit and 6-qubit states, Tapiador {\it et al} \cite{tapiador09} conducted a simulated annealing algorithm optimization procedure, searching for states not only highly entangled but also algebraically simple. They successfully found maximally entangled 5 and 6-qubit states with very simple algebraic structure. For 7 and 8-qubit systems, they also discovered highly entangled states with rather simple structure.

The simplicity of the structures of the states mentioned above means \cite{tapiador09} that the coefficients of the states with respect to the standard basis of the multi-qubit system are nice to write: only a sparse subset of the standard basis vectors have nonzero coefficients, and the coefficients are taken from a finite set of simple numbers ($\pm 1$, $\pm i$, ...). This is a natural requirement, since it is harder to derive quantum information protocols based on states lacking simple structures.

However, it is even more natural to demand that the maximally (highly) entangled states can be conveniently created by simple quantum circuits. If a maximally entangled state can be achieved by a sequence of simple quantum gates, it may be considered to have a certain conceptual simplicity, as well as (perhaps more importantly) a superior practical realizability.

Here we report research in which the genetic algorithm is used to search for quantum circuits producing maximally entangled multi-qubit \emph{pure} states numerically.  Using the negativity entanglement measure, we can construct simple quantum circuits for maximally or highly entangled states with up to 6 qubits, a method which, among other results, has led us to the simplest known maximally entangled 5 and 6-qubit quantum circuits.

Whether the GA is more effective than other optimization techniques at finding maximally entangled states for larger $n$ remains unknown, as the initial computational experiments reported here relied upon a GA implementation not suitable for highly scalable computation; so exploration of this question remains for future work.  However, our results do exemplify one significant advantage of our method over previous ones: it allows very simple construction of quantum circuits based on the quantum states found.

Section 2 reviews the entanglement measure used in the paper.
Section 3 reviews some relevant prior results on maximally entangled quantum systems.
Section 4 describes our use of genetic algorithms, including the details of the encoding scheme and the fitness function.  Section 5 presents results; and Section 6 gives conclusions and discussion.

\section{Measuring Entanglement}

We now review the  \emph{negativity} \cite{zyczkowski98,vidal02,kendon02}  entanglement measure which we use as the fitness function for our GA.  This measure has been shown computationally tractable in several previous works \cite{brown05,borras07,tapiador09}.

The negativity measure is closely related to the PPT \cite{peres96,horodecki97} (positive partial transpose) test, which states that positivity of the partial transpose of the density matrix of a bipartite state is a \emph{necessary} condition for the state to be separable.
Thus, an inseparable (entangled) state is characterized by non-vanishing \emph{negativity}, where the latter is defined as the sum of all the negative eigenvalues of the partial transpose of the density matrix.  It is convenient to define the entanglement $E_{\rm N}$ to be the negative of the negativity, i.e. $E_{\rm N} = - {\rm negativity}$. The larger $E_{\rm N}$, the more entangled the state. The maximally entangled 2-qubit states are the Bell states having
$E_{\rm N} = 0.5$.

It is straightforward to generalize the definition of negativity to multipartite states. In this paper, we consider systems with a fixed number of qubits, where each qubit constitutes one part of the whole system. Therefore, an $n$-partite state just means an $n$-qubit state.  For an $n$-qubit system, there are $C_n^0 + C_n^1 + ... + C_n^n = 2^n$ possible cuts (partitions). Each cut corresponds to a possible partial transpose. However, since each cut has an equivalent cut (e.g. the two cuts $\{0, 1\}$
\footnote{We label the qubits for an $n$-qubit system by the integers $0, 1, ..., n-1.$}
and $\{2, 3\}$ for a 4-qubit system are equivalent), and we do not need to consider the trivial partial transpose which does nothing, there are only $2^n/2 - 1 = 2^{n - 1} - 1$ nonequivalent cuts. The negativity for an $n$-qubit state is defined to be the sum of the negativity for the $2^{n - 1} - 1$ partial transposes and the entanglement $E_{\rm N}$ is still defined to be the negative of the negativity.  There are upper bounds of $E_{\rm N}$ for $n$-qubit states, which can be derived \cite{borras07} by considering a hypothetical $n$-qubit state whose marginal density matrices are all completely mixed. In this case, each $n$-cut partial transpose contributes an amount of $(2^{n} - 1)/2$ to $E_{\rm N}$. For example, the upper bounds of $E_{\rm N}$ for 3, 4, 5, and 6-qubit systems can be calculated to be  1.5, 6.5, 17.5, and 60.5 respectively. See Table 1.

\begin{table}[h]
\begin{center}
\begin{tabular}{|c||c|c|c|c|c|}
  \hline
  $n$  & 3 & 4 & 5 & 6 \\
  \hline
  1-cuts      & 3 & 4 & 5  & 6  \\
  \hline
  2-cuts      & 0 & 3 & 10 & 15 \\
  \hline
  3-cuts      & 0 & 0 & 0  & 10  \\
  \hline
  total cuts  & 3 & 7 & 15 & 31 \\
  \hline
  $E_{\rm N}$           & $0.5 \times 3 $
                        & $0.5 \times 4  $
                        & $0.5 \times 5 $
                        & $0.5 \times 6 + 1.5 \times 15  $\\
                        & $=1.5$
                        & $+ 1.5 \times 3 = 3.5$
                        & $+ 1.5 \times 10 = 17.5$
                        & $+ 3.5 \times 10 = 60.5$\\

\hline
\end{tabular}
\end{center}
\caption{Number of cuts for $n$-qubit systems and the calculation of the (hypothetical) maximal entanglement.}
\label{1}
\end{table}

\section{Prior Results on Maximally Entangled Multi-qubit Systems}

\subsection{Conventions for States, Gates and Circuits}

We use the following conventions for describing quantum states, gates, and circuits.

The $2^n$ computational basis of the $n$-qubit states is taken to be
$\{ |q_{n-1}\rangle |q_{n-2}\rangle \cdots |q_{0}\rangle \\
\equiv  |q_{n-1} q_{n-2} \cdots q_{0} \rangle \}$
where $q_i = 0$ or 1.
A general $n$-qubit state is expressed as a superposition of these basis states as
$|\Psi_n\rangle = \sum_{q_{n-1}q_{n-2}...q_{0}} c_{q_{n-1}q_{n-2} \cdots q_{0}}
|q_{n-1}q_{n-2} \cdots q_{0}\rangle$,
where the coefficients  $c_{q_{n-1}q_{n-2}\cdots q_{0}}$ are complex numbers such that the normalization condition
$\sum_{q_{n-1}q_{n-2}\cdots q_{0}} |c_{q_{n-1}q_{n-2}\cdots q_{0}}|^2 = 1$
for the state $|\Psi_n\rangle$ is satisfied. We also denote the four Bell basis states as
$|\psi^{\pm}\rangle = (|00\rangle \pm |11\rangle)/\sqrt{2}$ and
$|\phi^{\pm}\rangle = (|01\rangle \pm |10\rangle)/\sqrt{2}$.

The quantum gates acting on a general $n$-qubit state may be represented by $2^n$ dimensional matrices in the above basis. Elementary quantum gates can be conveniently expressed as tensor products of small matrices and appropriate identity matrices. For example, we use the notation $H(i)$ to represent a Hadamard gate acting on the $i$th qubit and use $\textmd{CNOT}(i, j)$ to represent a CNOT gate acting on the $i$th and the $j$th qubits, which are taken to be the control and the target qubits respectively.

A quantum circuit of \emph{size} $N$ will be represented by a string of $N$ elementary gates in the form
$\textmd{Circuit} =
\textmd{Gate}(N-1)\textmd{Gate}(N-2)\cdots \textmd{Gate}(1)\textmd{Gate}(0)$,
where Gate(0) acts first and Gate($N-1$) acts last. This convention is chosen to respect the convention of matrix multiplication -- which is opposite to the convention of the pictorial representations of quantum circuits, in which the gate at the leftmost acts first.

\subsection{4-qubit States}

Higuchi and Sudbery \cite{higuchi00} have proved that there is no 4-qubit pure state with all its marginal density matrices completely mixed, which means that the hypothetical maximal entanglement
($E_{\rm N}$ = 6.5)
is unreachable. Using variational methods, they found a highly entangled state
$|\textmd{HS}4\rangle$
with
$E_{\rm N} = 6.0981$,
\begin{equation}
\label{HS4}
|\textmd{HS}4\rangle = \frac{1}{\sqrt{6}}\big(
|1100\rangle + |0011\rangle + \omega(|1001\rangle + |0110\rangle)
+ \omega^2(|1010\rangle + |0101\rangle)
\big)
\end{equation}
where
$\omega = -1/2 + i\sqrt{3}/2$
is the third root of unity. This state is known to be a local maximum \cite{brierley07} and is also conjectured to be a global maximum \cite{higuchi00} according to the von Neumann entropy measure.

\subsection{5-qubit States}

Maximally entangled 5-qubit states have been discovered via previous experiments with numerical search algorithms \cite{brown05,borras07,tapiador09}. These states all have the same entanglement feature as the following state,
\begin{equation}\label{BSSB5}
|\textmd{BSSB}5\rangle = \frac{1}{2}(
|001\rangle|\phi^{-}\rangle + |010\rangle|\psi^{-}\rangle +
|100\rangle|\phi^{+}\rangle + |111\rangle|\psi^{+}\rangle).
\end{equation}
The entanglement distribution of this state among different cuts can be expressed as $E_{\rm N} = 5 \times 0.5 + 10 \times 1.5 = 17.5$.

Muralidharan and Panigrahi \cite{muralidharan08} found various quantum information
applications of this state and outlined the procedure of a possible physical realization of this state. They proposed to create this state by using 2 $H$s and 3 CNOTs followed by a 32 dimensional matrix with prescribed matrix elements, but the decomposition of this matrix into elementary quantum gates was not provided. The advantage of our algorithm, to be reviewed below, is that we can find not only maximally entangled 5-qubit states, but also the corresponding quantum circuits creating the states.

\subsection{6-qubit states}

By running a hill climbing algorithm, Borras {\it et al} \cite{borras07} discovered a maximally entangled 6-qubit state with 32 non-vanishing coefficients, which was used by Choudhury {\it et al} \cite{choudhury09} for various quantum information applications.
Maximally entangled 6-qubit states with 16 nonzero coefficients were also discovered by Tapiador {\it et al} \cite{tapiador09} using the simulated annealing algorithm.
Using our algorithm, we can find both of these kinds of states and the corresponding circuits creating them.

\section{Applying the Genetic Algorithm}

We now describe our use of a genetic algorithm to evolve quantum circuits with maximal possible entanglement according to the negativity measure.

\subsection{Brief Review of Genetic Algorithms}

The genetic algorithm (GA) \cite{goldberg89} is an optimization algorithm encapsulating the basic ideas of biological evolutionary theory (mutation, combinatory sexual reproduction, and differential reproduction in a population based on fitness).  GAs and other related ``evolutionary algorithms" have been applied to quantum information and computation before, e.g. to the automated generation of quantum algorithms; see \cite{spector04} and \cite{gepp09} for reviews.

The basic idea of the GA is rather simple. Initially, a starting population of individuals (constituting possible solutions to the problem at hand) is generated randomly.   In our problem, an individual is a quantum circuit which can be used to create a new quantum state from an initial state. In the GA broadly conceived, an individual has two aspects. On one hand, it has a genotype, which is a sequence of genes (also called a chromosome); on the other hand, it has a phenotype, which determines its fitness (the quality of the solution). The fitness of an individual is calculated using the fitness function which, in our problem is just the entanglement $E_{\rm N}$.  The translations from the phenotype to its genotype and vice versa are called encoding and decoding respectively. The implementation of these ideas in the context of evolving maximally entangled quantum circuits will be detailed in the next subsection.

After an initial population is created, the GA starts the evolutionary loop, which consists of the following steps:

\begin{itemize}
\item evaluate the fitness of each individual in the current population (generation)
\item select individuals based on the fitness levels to be parents of the next generation
\item generate the next generation through genetic operations, namely
\begin{itemize}
\item crossover, a binary operation that takes in two chromosomes and outputs a new one combining genes from each of them
\item mutation, which alters a certain percentage of genes in a chromosome, usually randomly (sometimes drawn from a particular probability distribution)
\end{itemize}
\end{itemize}

At each generation, the best solution (the elite) in the population is always retained for the next generation, until  better solutions are found. This loop terminates when sufficiently optimized individuals have been found or other predetermined termination criteria (such as a maximum number of generations) are met.

\subsection{Encoding of the quantum circuits}

Our aim, in the present research, is to find simple quantum circuits which can create maximally entangled states using the GA.  To do this, we first encode all possible quantum circuits (solutions) into chromosomes on which the genetic operators can act.  There are many feasible encoding schemes from which we can choose. Since our problem is discrete in nature, we can use binary or integer representations for the chromosomes, and we have chosen the latter.

A quantum circuit is a sequence of elementary quantum gates acting on a number of qubits. Any quantum circuit (any unitary matrix) can be approximated by a set of universal quantum gates. One set of such universal gates consists of the Hadamard gate $H$, the $\pi/8$ gate $T$, and the controlled-NOT gate CNOT. Other useful elementary quantum gates include the three Pauli gates $X$, $Y$, and $Z$, the phase
gate $S$, and the controlled-$Z$ gate \cite{nielsen00}. Not all the gates are needed to construct a quantum circuit for a given $n$-qubit state. For example, the Bell states can be prepared using only a Hadamard gate followed by a CNOT gate. Therefore, in our computational experiments we have tried various sets of elementary quantum gates to generate states of a given number of qubits.

The elementary quantum gates mentioned above are all 1 and 2-qubit gates.
When acting on a multi-qubit state, each of these gates can act (nontrivially) on different qubit(s). Take the 6-qubit states as an example. If we use $H$ and CNOT as our elementary quantum gates, there will be 6 different $H$s acting on each one of the 6 qubits and 30 CNOTs acting on different (ordered) pairs of qubits. We can arrange all 30 elementary quantum gates into an array, in which each one of the 30 elementary gates is specified by an integer between 0 and 35. In this way, a quantum circuit for a 6-qubit state using say, 10 elementary gates is represented by an array of 10 integers, each of which ranges from 0 to 35. This array of integers is a chromosome, and each integer corresponds to a gene. Since there are 10 genes and each gene can take 36 different values, the size of the search space (the number of all possible solutions) is $36^{10} \simeq 10^{15}$.  The size of this search space renders brute force search infeasible on current computers; but using the GA, we can find an optimized solution using only hundreds to thousands of evaluations of the fitness function.

\subsection{Evaluation of the fitness function}
While the chromosome is the object which the genetic operators act on, the fitness is the driving force for GA to evolve the population of individuals. The fitness function accepts a chromosome as its input and returns a scalar (a real number) as its output. The evaluation of the fitness function for our problem takes on the following steps:

\begin{itemize}
\item translate the chromosome (array of integers) into the corresponding quantum circuit (sequence of elementary quantum gates)
\item obtain the target state by acting on the initial state (taken as $|00\cdots0\rangle$) with the quantum circuit
\item calculate the fitness $E_{\rm N}$ of the state obtained in the last step using the method given above
\end{itemize}

\section{Results}

We now report the results we have obtained via implementing the above ideas using a GA implemented using the genetic algorithm toolbox of MATLAB \cite{matlab}, for the case of 3, 4, 5, and 6-qubit states.

\begin{figure}
\resizebox{1\columnwidth}{!}{%
  \includegraphics{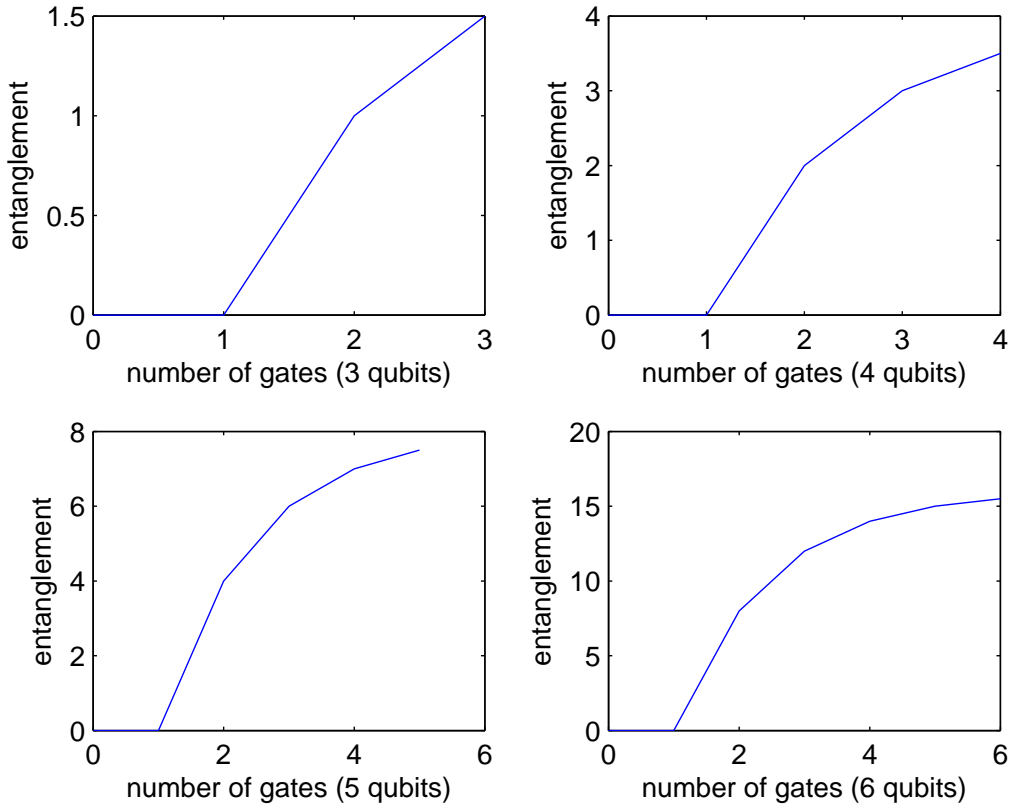}
}
\caption{The entanglement generated by the quantum gates at each step of the quantum circuits for the 3, 4, 5, and 6-qubit GHZ states.}
\label{fig:GHZstates}
\end{figure}

\subsection{3 Qubits}

For the 3 qubit case, the GA easily finds the GHZ state, as expected.
For example, the GHZ state
$|\textmd{GHZ}3\rangle = (|000\rangle + |111\rangle)/\sqrt{2}$
is prepared by the following circuit with only 3 elementary gates acting on the initial state $|000\rangle$,
\begin{equation}
\label{circutGHZ3}
\textmd{Circuit}_{\rm GHZ3} =
\textmd{CNOT}(2,0)\textmd{CNOT}(2,1) H(2).
\end{equation}
The entanglement is generated in the following way. Firstly, $H(2)$ change the initial state
$|000\rangle$
to a superposition of
$|000\rangle$ and $|100\rangle$
without generating any entanglement. Then, CNOT(2,1) prepares a Bell state
$(|00\rangle + |11\rangle)|0\rangle/\sqrt{2}$
for the two qubits $q_1$ and $q_2$, generating an amount of entanglement
$E_{\rm N} = 1$.
Finally, CNOT(2,0) promotes the entanglement to the maximal value
$E_{\rm N} = 1.5$,
making the 3 qubits completely entangled. See Figure~\ref{fig:GHZstates}.

Generalizing this construction, the $n$-qubit GHZ state can be prepared by the following circuit with $n$ elementary quantum gates,
\begin{equation}
\label{circutGHZn}
\textmd{Circuit}_{{\rm GHZ}n} =
\textmd{CNOT}(n-1,0)\cdots \textmd{CNOT}(n-1,n-2) H(n-1).
\end{equation}
Again, the Hadamard gate prepares superpositions and the CNOT gates generate entanglements. The amount of entanglement generated by
$\textmd{CNOT}(n-1, m) (0\leq m \leq n-2)$
is $E_{\rm N} = 2^{m-1}$ and the total entanglement for the $n$-qubit GHZ state is $E_{\rm N} = (2^{n-1} - 1)/2$. See Figure~\ref{fig:GHZstates}.

\begin{figure}
\resizebox{1\columnwidth}{!}{%
  \includegraphics{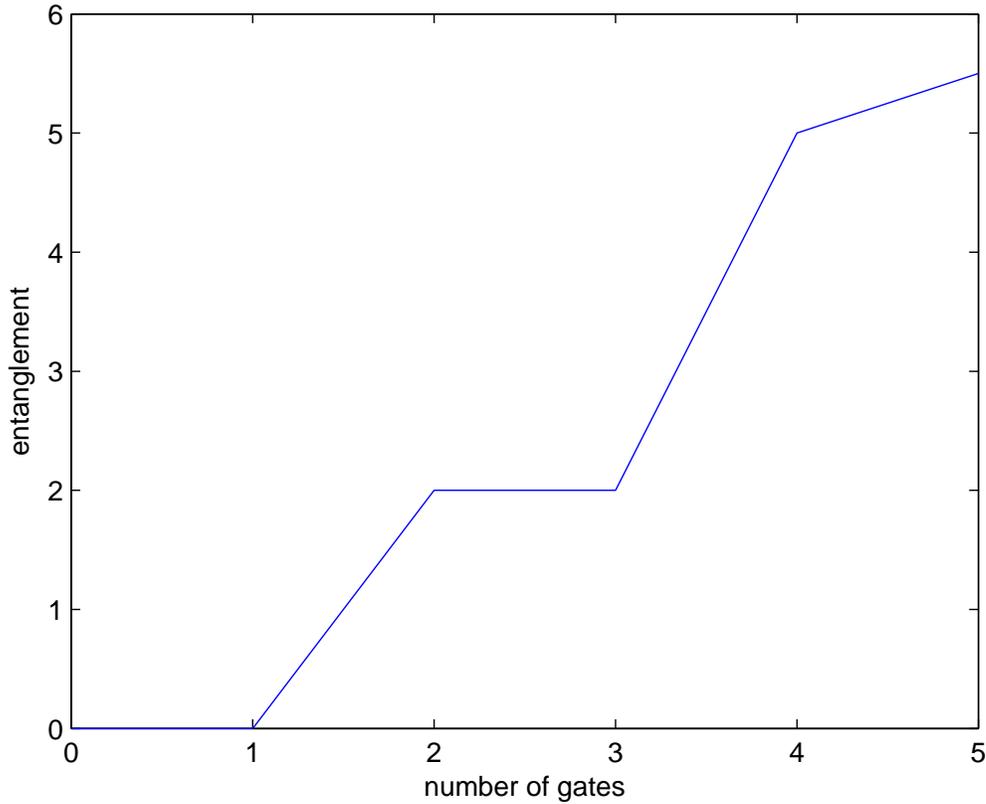}
}
\caption{The entanglement generated by the quantum circuit for $|\Psi_{4a(b)}\rangle$ at each step.}
\label{fig:4qubits}
\end{figure}

\subsection{4 Qubits}

For the 4 qubit case the result is more interesting.  Using $H$ and CNOT as the elementary quantum gates, the best solution found by the GA has
$E_{\rm N} = 5.5$
and the simplest quantum circuit we found is of size 5,
\begin{equation}\label{circut5}
\textmd{Circuit}_{4a} =
\textmd{CNOT}(3,1)\textmd{CNOT}(3,0) H(3)\textmd{CNOT}(2,1) H(2).
\end{equation}
The state created by this quantum circuit has the following form,
\begin{equation}
\label{Psi4a}
\begin{array}{rl}
|\Psi_{4a}\rangle = &\frac{1}{2}\big(
|0000\rangle + |0110\rangle + |1011\rangle + |1101\rangle \big) \\
= & \frac{1}{\sqrt{2}}\big(
|0\rangle |\psi^{+}\rangle |0\rangle +
|1\rangle |\phi^{+}\rangle |1\rangle
\big).
\end{array}
\end{equation}

There are 7 cuts for a 4-qubit system, 4 single-cuts and 3 two-cuts.
The marginal density matrices of $|\Psi_{4a}\rangle$ for the 4 single-cuts are all completely mixed. For 2 of the 3 two-index cuts, $\{0,1\}$ and $\{0,2\}$, the marginal density matrices are also completely mixed, but the marginal density matrix for the two-index cut $\{0,3\}$ is not. The entanglement distribution of this state can be expressed as $E_{\rm N} = 4 \times 0.5 + 2 \times 1.5 + 0.5 = 5.5$. We note that this state has the same entanglement feature as the one used by Wu and Zhang to show that not all 4-partite pure states  are GHZ reducible \cite{wu01}.

By a permutation of the qubits (This is natural, since all the qubits are equivalent and there is no privilege for any qubit \cite{gisin98}), this state can be brought to a nicer form,
\begin{equation}
\label{Psi4b}
|\Psi_{4b}\rangle =  \frac{1}{\sqrt{2}}\big(
|00\rangle |\psi^{+}\rangle +
|11\rangle |\phi^{+}\rangle
\big).
\end{equation}
The quantum circuit for the state $|\Psi_{4b}\rangle$ can be deduced from $\textmd{Circuit}_{4a}$ to be
\begin{equation}\label{circut5}
\textmd{Circuit}_{4b} =
\textmd{CNOT}(3,0)\textmd{CNOT}(3,2) H(3)\textmd{CNOT}(1,0) H(1).
\end{equation}

Although less entangled than the HS state, $\Psi_{4a(b)}$ has a simpler form and is much more entangled than an average 4-qubit state as can be seen from the entanglement distribution of for 4-qubit states \cite{borras07}. The entanglement generated by $\textmd{Circuit}_{4a(b)}$ is shown in Figure~\ref{fig:4qubits}.
Incidently, we note that by using a GA with continuous chromosome representation and searching the states only, the algorithm always converges to the same entanglement value as the HS state $E_{\rm N} = 60.981$. This conforms once again that the HS state is the global maximally entangled 4-qubit state.

We can interpret the quantum circuit $\textmd{Circuit}_{4a(b)}$ in the following way.
Firstly, two Bell pairs are prepared by two sets of Hadamard and CNOT gates acting on two pairs of qubits. The entanglement generated at this stage is $E_{\rm N} = 5$. Secondly, a CNOT gate with the control qubit choosing from one pair of Bell qubits and the target qubit choosing from the other pair of Bell qubits increases the entanglement to
$E_{\rm N} = 5.5$.

\begin{figure}
\resizebox{1\columnwidth}{!}{%
  \includegraphics{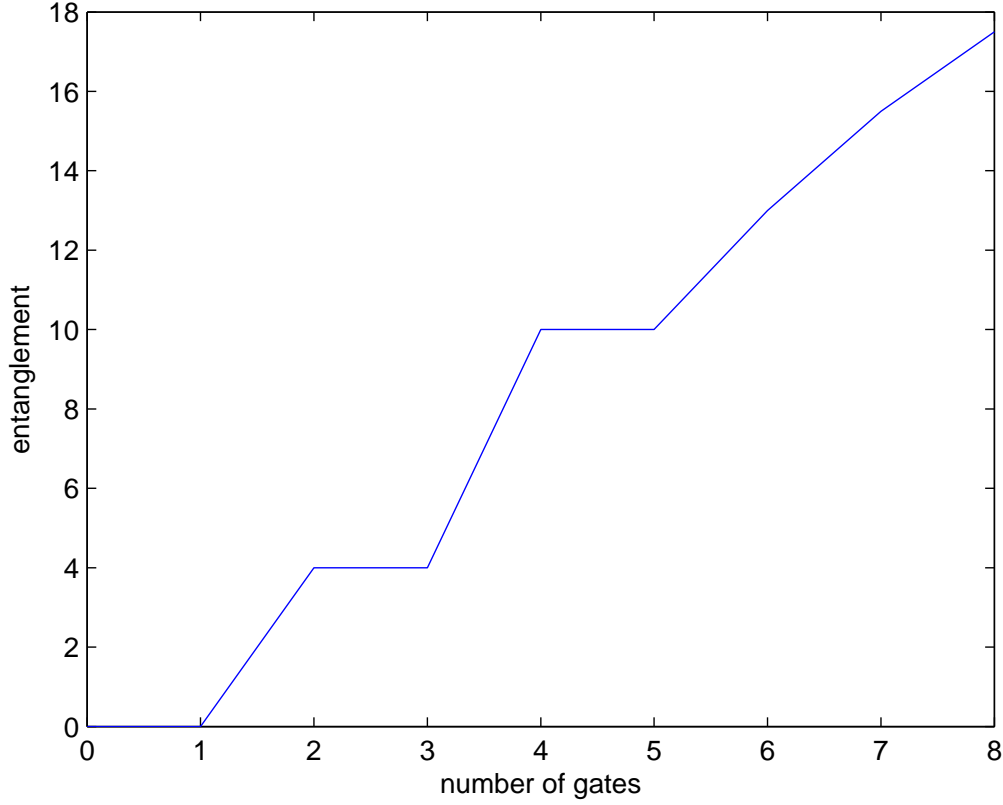}
}
\caption{The entanglement generated by the quantum circuit for $|\Psi_{5a(b)}\rangle$ at each step.}
\label{fig:5qubits}
\end{figure}

\subsection{5 Qubits}

We have tried various sets of elementary quantum gates in our GA experiments aimed at evolution of maximally entangled 5-qubit states.
It turns out that it suffices to use the set of elementary quantum gates consisting of 5 Hadamard gates $H(i)$ $(0\leq 4)$ and 20 CNOT gates $\textmd{CNOT}(i, j)$ $(0\leq i, j \leq 4; i \neq j)$.

Using these gates as the elementary quantum gates, a maximally entangled state similar to $|\textmd{BSSB}5\rangle$ can be easily found by the GA. As we increase the size of the quantum circuit, the maximal entanglement obtainable increases monotonically. When the size of the circuit equals to 8, the maximal entanglement $E_{\rm N} = 17.5$ is achieved.

One of the maximally entangled states discovered takes the following form,
\begin{equation}\label{Psi5a}
\begin{array}{rl}
|\Psi_{5a}\rangle = &
\frac{1}{\sqrt{8}}(|00000\rangle + |00111\rangle + |01011\rangle + |01100\rangle \\
& + |10010\rangle + |10101\rangle - |11001\rangle - |11110\rangle ) \\
= & \frac{1}{2} (
|0\rangle |\psi^{+}\rangle |00\rangle +
|0\rangle |\phi^{+}\rangle |11\rangle +
|1\rangle |\phi^{-}\rangle |01\rangle +
|1\rangle |\psi^{-}\rangle |10\rangle ),
\end{array}
\end{equation}
which is created from the following circuit acting on the initial state $|00000\rangle$,
\begin{equation}\label{circut5a}
\begin{array}{rl}
\textmd{Circuit}_{5a} = &
\textmd{CNOT}(3,2) \textmd{CNOT}(4,1) \textmd{CNOT}(1,0)\\
& H(4)\textmd{CNOT}(4,3) H(4)\textmd{CNOT}(2,1) H(2).
\end{array}
\end{equation}
By a permutation of the qubits, we can bring the state $|\Psi_{5a}\rangle$ to a nicer form,
\begin{equation}\label{Psi5b}
|\Psi_{5b}\rangle =  \frac{1}{2} (
|000\rangle |\psi^{+}\rangle  +
|011\rangle |\phi^{+}\rangle  +
|101\rangle |\phi^{-}\rangle  +
|110\rangle |\psi^{-}\rangle  ),
\end{equation}
which can be created from the following circuit acting on the initial state $|00000\rangle$,
\begin{equation}\label{circut5b}
\begin{array}{rl}
\textmd{Circuit}_{5b} = &
\textmd{CNOT}(1,0) \textmd{CNOT}(4,3) \textmd{CNOT}(3,2)\\
& H(4)\textmd{CNOT}(4,1) H(4) \textmd{CNOT}(0,3) H(0).
\end{array}
\end{equation}

The entanglement generated by $\textmd{Circuit}_{5a(b)}$ is shown in Figure~\ref{fig:5qubits}. In this case, two Bell pairs are firstly created, followed by a Hadamard gate and 3 CNOT gates. We note that at least 3 Hadamard gates are needed to generate maximal entanglement, which means that maximally entangled 5-qubit states have at least 8 non-vanishing coefficients.

\begin{figure}
\resizebox{1\columnwidth}{!}{%
  \includegraphics{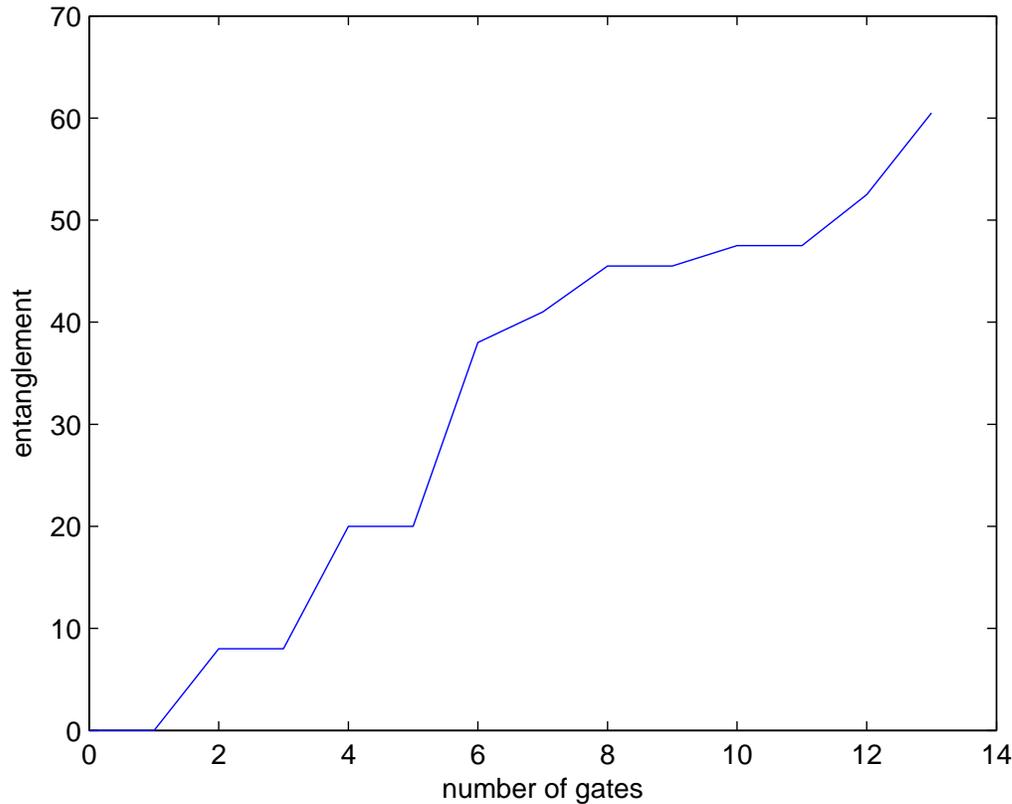}
}
\caption{The entanglement generated by the quantum circuit for $|\Psi_{6}\rangle$ at each step.}
\label{fig:6qubits}
\end{figure}

\subsection{6 Qubits}

Similarly, we can find a quantum circuit producing maximal entanglement for 6-qubit system easily using the GA.

As with the 5-qubit case, quantum circuits for maximally entangled 6-qubit states can be found by the GA using only the Hadamard and the CONT gates. Therefore, the set of elementary quantum gates which can generate maximal entanglement for 6-qubit states consists of 6 Hadamard gates $H(i)$ $(0\leq 5)$ and 30 CNOT gates $\textmd{CNOT}(i, j)$ $(0\leq i, j \leq 5; i \neq j)$.

The shortest circuit creating maximal entanglement found by our algorithm has 13 gates, 5 $H$s and 8 CNOTs. A typical maximally entangled 6-qubit state with 32 non-vanishing coefficients in this case is,
\begin{equation}
\label{Psi6a}
\begin{array}{rl}
|\Psi_{6a}\rangle = &
\frac{1}{\sqrt{32}}
(   |000000\rangle + |000001\rangle + |000010\rangle - |000011\rangle \\
& - |001100\rangle + |001101\rangle + |001110\rangle + |001111\rangle \\
& + |010100\rangle + |010101\rangle - |010110\rangle + |010111\rangle \\
& + |011000\rangle - |011001\rangle + |011010\rangle + |011011\rangle \\
& + |100100\rangle - |100101\rangle + |100110\rangle + |100111\rangle \\
& + |101000\rangle + |101001\rangle - |101010\rangle + |101011\rangle \\
& + |110000\rangle - |110001\rangle - |110010\rangle - |110011\rangle \\
& - |111100\rangle - |111101\rangle - |111110\rangle + |111111\rangle )\\
= & \frac{1}{4}
(  (|0000\rangle - |1111\rangle) ( |\psi^{-}\rangle + |\phi^{+}\rangle)\\
&+ (|0011\rangle - |1100\rangle) (-|\psi^{-}\rangle + |\phi^{+}\rangle)\\
&+ (|0101\rangle + |1010\rangle) ( |\psi^{+}\rangle + |\phi^{-}\rangle)\\
&+ (|0110\rangle + |1001\rangle) ( |\psi^{+}\rangle - |\phi^{-}\rangle)).
\end{array}
\end{equation}
which is created by the following circuit of size 13 from the initial state $|000000\rangle$,
\begin{equation}
\label{circut6a}
\begin{array}{rl}
\textmd{Circuit}_{6a} =
&\textmd{CNOT}(2,1)\textmd{CNOT}(4,1)H(1)\textmd{CNOT}(4,3)\\ &H(4)\textmd{CNOT}(5,2)\textmd{CNOT}(3,0)\textmd{CNOT}(5,4)\\
&H(5)\textmd{CNOT}(3,2)H(3)\textmd{CNOT}(1,0)H(1).
\end{array}
\end{equation}

The state $|\Psi_{6a}\rangle$ is maximally entangled, $E_{\rm N} = 6 \times 0.5 + 15 \times 1.5 + 10 \times 3.5= 60.5$. See Figure~\ref{fig:6qubits} for the generation of entanglement of $\textmd{Circuit}_{6a}$. In this case, 3 Bell pairs are created before the action of the other 2 Hadamard and 5 CNOT gates.
We note that $\Psi_{6a}$ has a more compact form than the maximally entangled 6-qubit state discovered by Borras {\it et al} \cite{borras07,choudhury09}. Moreover, by applying a Hadamard gate say, $H(0)$, to $|\Psi_{6a}\rangle$, we can get a maximally entangled 6-qubit state with only 16 nonvanishing coefficients,
\begin{equation}\label{Psi6b}
\begin{array}{rl}
|\Psi_{6b}\rangle = &
 \frac{1}{\sqrt{8}}
(       (|0000\rangle - |1111\rangle) |\psi^+\rangle
      + (-|0011\rangle + |1100\rangle) |\phi^-\rangle\\
    & + (|0101\rangle + |1010\rangle) |\psi^-\rangle
      + (|0110\rangle + |1001\rangle) |\phi^+\rangle),
\end{array}
\end{equation}
which is essentially equivalent to the state found by Tapiador {\it et al} \cite{tapiador09}.

Our experiments also suggest that 16 is the minimal number of the nonvanishing coefficients for the maximally entangled 6-qubit states created from the Hadamard and CNOT gates.

\section{Conclusions and Future Work}

In this paper, we have applied the GA to the automated design of quantum circuits for maximally  entangled states. Highly entangled 4-qubit states and maximally entangled 5 and 6-qubit states have been found, including the corresponding quantum circuits creating them.

The quantum circuits for the maximally (highly) entangled states are very simple, using only 3, 5, 8, and 13 elementary gates for 3, 4, 5, and 6-qubit systems respectively.  The elementary gates used only include the Hadamard and the CNOT gates. The Hadamard gates play the role of generating \emph{quantum parallelism} (generating non-vanishing coefficients of the state) and the CNOT gates play the genuine role of \emph{entanglers}. Our numerical results suggest that maximally entangled multi-qubit states are more-than-two superpositions of the computational basis states. The minimal numbers of nonzero coefficients for maximally entangled 5 and 6-qubit states are found to be 8 and 16 respectively. This explains partly why the GHZ states are not the maximally entangled states for 4 and more qubit systems.

We note that the construction of quantum circuits for maximally entangled states is similar to the construction of \emph{perfect entanglers} \cite{kraus01} from imperfect ones. A perfect 2-qubit entangler is a quantum gate which can generate maximal entanglement from an unentangled state. The CNOT gate is a perfect entangler for 2-qubit states, but for an $n$-qubit ($n\geq3$) system, the CNOT gates are not perfect entanglers since each one of them generates only a (small) part of the whole entanglement of the $n$-qubit system. In this sense, the quantum circuits for maximally entangled states presented in this paper can be seen as perfect entanglers for multi-qubit systems.

The performance of our algorithm appears similar to that of previous approaches using hill climbing or simulated annealing. The running time for a single fitness evaluation ($E_{\rm N}$) increases exponentially with the number of qubits. The number of evaluations of the fitness function needed before finding an optimal solution also increases, though the scaling here is less clear.  A typical run of the algorithm for the 6-qubit case on a common contemporary laptop takes a few minutes, evaluating the fitness function several thousand times.

Envisioned future work includes experimenting with a more scalable C++ GA implementation, which may allow search through spaces associated with 7 and more qubit systems.  We also intend to experiment with alternative entanglement measures including the EMM (the entanglement measure based on minors), which is hypothesized to be more computationally tractable than eigenvalue based entanglement measures \cite{long09}.  Use of more advanced evolutionary algorithms such as the Bayesian Optimization Algorithm \cite{pelikan05} may also be helpful; or the use of estimation-of-distribution algorithms such as MOSES \cite{looks06} that act directly on circuit space rather than utilizing an integer genome.

\ack
This work is supported by National Natural Science Foundation of China under grant Nos 10535010, 10675090, 10775068, and 10735010.

\end{document}